\documentclass[12pt]{article}
\pdfoutput=1
\usepackage{amsmath,amssymb}
\usepackage{graphicx}
\usepackage[pdftex]{hyperref}
\DeclareGraphicsExtensions{.eps,.bmp,.wmf,.jpg,.pdf}
\numberwithin{equation}{section}
\def\be{\begin{equation}}
\def\ee{\end{equation}}

\def\bea{\begin{eqnarray}}
\def\eea{\end{eqnarray}}

\title{Infrared cut-off proposal for the Holographic density}
\author{L.N. Granda\thanks{ngranda@univalle.edu.co} \, and\  A. Oliveros\thanks{alexogar@univalle.edu.co}\\
Department of Physics, Universidad del Valle\\ A.A. 25360, Cali,
Colombia} 
\date{}
\begin{document}
\maketitle

\begin{abstract}
\noindent We propose an infrared cut-off for the holographic the dark-energy, which besides the square of the Hubble scale also contains the time derivative of the Hubble scale. This avoids the problem of causality which appears using the event horizon area as the cut-off, and solves the coincidence problem. 

\noindent \it{PACS: 98.80.-k, 95.36.+x}\\
\noindent \it{Keywords: Holography; dark energy; redshift}

\end{abstract}

\section{Introduction}
\noindent 
Recent astrophysical data from distant Ia supernovae observations \cite{SN},\cite{riess} show that the current Universe is not only expanding, but also it is accelerating due to some kind of  negative-pressure form of matter known as dark energy (\cite{copeland},\cite{turner}). The simplest candidate for dark energy is the cosmological constant \cite{weinberg}, conventionally associated with the energy of the vacuum with constant energy density and pressure, and an equation of state $w=-1$. The present observational data favor an equation of state for the dark energy with parameter very close to that of the cosmological constant. The next simple model proposed for dark energy is the quintessence (see \cite{copeland1}, \cite{caldwell}, \cite{zlatev}), a dynamical scalar field which slowly rolls down in a flat enough potential. The equation of state for a spatially homogeneous quintessence scalar field satisfies $w>-1$ and therefore can produce accelerated expansion. This field is taken to be extremely light which is compatible with its homogeneity and avoids the problem with the initial conditions. 

More exotic models proposed to explain the nature of the dark energy, are related with K-essence models based on scalar field with non-standard kinetic term \cite{armendariz},\cite{chiba}; string theory fundamental scalars known as tachyon \cite{padmana} and dilaton \cite{gasperini}; scalar field with negative kinetic energy, which provides a solution known as phantom dark energy \cite{caldwell1}, and Chaplygin gas \cite{kamen} among others (for a review on above mentioned and other approaches, see \cite{copeland}) . An alternative approach to dark energy is related to modified theory of gravity $f(R)$ (\cite{carroll},\cite{capozziello}, \cite{odintsov}, \cite{sergei1}), in which dark energy emerges from the modification of geometry. Of course this modifications should pass precise solar system tests, which leads to the necessity of fine tunning in the additional terms, and this significantly restricts the possible form of the $f(R)$ gravity. 

Recent studies of black holes and string theories may provide a new alternative to the solution of the dark energy problem, known as the holographic principle (\cite{beckenstein, thooft, bousso, cohen}). This principle emerges as a new paradigm in quantum gravity and was first put forward by t' Hooft \cite{thooft} in the context of black hole physics and later extended by Susskind \cite{susskind} to string theory. Acording to the holographic principle, the entropy of a system scales not with it's volume, but with it's surface area (\cite{bousso, susskind}). In other words, the degrees of freedom of a spatial region reside not in the bulk but only at the boundary of the region and the number of degrees of freedom per Planck area is no greater than unity.
Applied to cosmology, Fischler and Susskind \cite{fischler} have proposed a  version of the holographic principle: at any time during cosmological evolution, the gravitational entropy within a closed surface should be always larger than the particle entropy that passes through the past light-cone of that surface. In the case of the standard big-bang cosmology, they have found that only open or flat universe but not closed one is compatible with the cosmological holographic principle, provided one makes certain assumptions on the initial big-bang singularity.\\

In the work \cite{cohen}, it was suggested that in quantum field theory a short distance cut-off is related to a long distance cut-off due to the limit set by formation of a black hole, namely, if is the quantum zero-point energy density caused by a short distance cut-off, the total energy in a region of size L should not exceed the mass of a black hole of the same size, thus $L^3\rho_\Lambda\leq LM_p^2$. The largest $L$ allowed is the one saturating this inequality, thus
\begin{equation}\label{eq1}
\rho_\Lambda=3c^2M_p^2L^{-2}
\end{equation}

In the context of the dark energy problem, initially the holographic principle  proposes that essentially the unknown vacuum energy density $\rho_{\Lambda}$ is proportional to the square of the Hubble scale $\rho_{\Lambda}\propto H^2$. This in principle solves the fine tunning problem, but the equation of state is zero and does not contribute to the present accelerated expansion. As was shown in work \cite{li}, using the particle horizon as the length scale gives an equation of state parameter higher than $-1/3$, which neither explain the present acceleration, but the future event horizon gives the desired acceleration regime, although this model faces the causality problem.

For purely dimensional reasons we propose a new infrared cut-off for the holographic density which includes time derivative of the Hubble parameter, and in this paper we study the fitting of this model with the current observational data. In favor of this new term we can say that the underlying origin of the holographic dark energy is still unknown and that the new term is contained in the expression for the Ricci scalar which scales as $L^{-2}$ (a model with holographic dark energy proportional to the Ricci scalar was proposed in \cite{gao}). So, we propose a holographic density  of the form $\rho\approx\alpha H^2+\beta\dot{H}$.

\section{The Model}

Let us start with the following  holographic dark energy density:
\begin{equation}\label{eq2}
\rho_\Lambda=3\left(\alpha H^2+\beta\dot{H}\right)
\end{equation}

\noindent where $\alpha$ and $\beta$ are constants to be determined and $H=\dot{a}/a$ is the Hubble parameter. The usual Friedmann equation is
\begin{equation}\label{eq3}
H^2=\frac{1}{3}\left(\rho_m+\rho_r+\rho_\Lambda\right)
\end{equation}
where we have taken $8\pi G=1$ and $\rho_{m}$, $\rho_{r}$ terms are the contributions of non-relativistic matter and radiation, respectively.
\noindent Setting $x=\ln{a}$, we can rewrite the Friedmann equation as follows
\begin{equation}\label{eq4}
H^2=\frac{1}{3}\left(\rho_{m0}e^{-3x}+\rho_{r0}e^{-4x}\right)+\alpha H^2+\frac{\beta}{2}\frac{dH^2}{dx}
\end{equation}

\noindent Introducing the scaled Hubble expansion rate $\tilde{H}=H/H_0$, where $H_0$ is the present value of the Hubble constant (for $x=0$), the above Friedman equation becomes
\begin{equation}\label{eq5}
\tilde{H}^2=\Omega_{m0}e^{-3x}+\Omega_{r0}e^{-4x}+\alpha\tilde{H}^2
+\frac{\beta}{2}\frac{d\tilde{H}^2}{dx}
\end{equation}
where $\Omega_{m0}=\rho_{m0}/3H_0^2$ and $\Omega_{r0}=\rho_{r0}/3H_0^2$ are the current density parameters
of non-relativistic matter and radiation. The last two terms in the above equation, valuated at $x=0$, represent the current holographic dark energy density parameter $\Omega_{\Lambda 0}$. These densities satisfy the constraint from Eqs. \ref{eq3}, \ref{eq5} $\Omega_{m0}+
\Omega_{r0}+\Omega_{\Lambda 0}=1$. Solving the equation (\ref{eq5}), we obtain
\begin{equation}\label{eq6}
\begin{aligned}
\tilde{H}^2=&\Omega_{m0}e^{-3x}+\Omega_{r0}e^{-4x}+\frac{3\beta-2\alpha}{2\alpha-3\beta-2}\Omega_{m0}e^{-3x}\\
&+\frac{2\beta-\alpha}{\alpha-2\beta-1}\Omega_{r0}e^{-4x}+Ce^{-2 x(\alpha-1)/\beta}
\end{aligned}
\end{equation}
where $C$ is an integration constant and  the last three terms give the
scaled dark energy density, which we will represent as $\tilde{\rho}_{\Lambda}=\frac{\rho_{\Lambda}}{3H_0^2}$:
\begin{equation}\label{eq7}
\tilde{\rho}_{\Lambda}=\frac{3\beta-2\alpha}{2\alpha-3\beta-2}\Omega_{m0}e^{-3x}\\
+\frac{2\beta-\alpha}{\alpha-2\beta-1}\Omega_{r0}e^{-4x}+Ce^{-2 x(\alpha-1)/\beta}
\end{equation}

\noindent Substituting the expression for $\tilde{\rho}_{\Lambda}$ into the energy conservation equation,
\begin{equation}\label{eq8}
\tilde{p}_{\Lambda}=-\tilde{\rho}_{\Lambda}-\frac{1}{3}\frac{d\tilde{\rho}_{\Lambda}}{dx}
\end{equation}
we obtain the dark energy pressure
\begin{equation}\label{eq9}
\tilde{p}_{\Lambda}=\frac{2\alpha-3\beta-2}{3\beta}\,Ce^{-2 x(\alpha-1)/\beta}+\frac{2\beta-\alpha}{3(\alpha-2\beta-1)}\,
\Omega_{r0}e^{-4x}
\end{equation}

\noindent There are three constants $\alpha$, $\beta$ and $C$ to be determined in the expressions (\ref{eq7})
and (\ref{eq9}). Considering the equation of state for the present epoch values of the density and pressure (i.e. at x=0) of the dark energy, 
$\tilde{p}_{\Lambda0}=\omega_0\Omega_{\Lambda0}$, we obtain (note that $\tilde{\rho}_{\Lambda0}=\Omega_{\Lambda0})$
\begin{equation}\label{eq10}
\begin{aligned}
C=&1+\frac{2\Omega_{m0}}{2(\Omega_{\Lambda 0}-1)+\beta(3\Omega_{m0}+4\Omega_{r0}+3(1+\omega_0)\Omega_{\Lambda 0}-3)}\\
&+\frac{2\Omega_{r0}}{2(\Omega_{\Lambda 0}-1)+\beta(3\Omega_{m0}+4\Omega_{r0}+3(1+\omega_0)\Omega_{\Lambda 0}-4)}
\end{aligned}
\end{equation}
and
\begin{equation}\label{eq11}
\alpha=\frac{1}{2}(2\Omega_{\Lambda 0}+\beta(3\Omega_{m0}+4\Omega_{r0}+3(1+\omega_0)\Omega_{\Lambda 0}))
\end{equation}
where the constants $C$ and $\alpha$ are given in terms of the constant $\beta$, which will be fixed by the behavior of the deceleration parameter versus the redshift $z$, adjusting the value of $\beta$ in order to obtain $z_T$ at which the deceleration parameter pases from the deceleration to acceleration regime \cite{copeland}.
The deceleration parameter is given by
\begin{equation}\label{eq12}
q=\frac{1}{2}+\frac{3\tilde{p}_{\Lambda}}{2(\tilde{\rho}_{\Lambda}+\tilde{\rho}_m)}
\end{equation}
where in what follows we despise the contribution from radiation, $p_m=0$ for dust matter, $\tilde{\rho}_m=\rho_m/3H_0^2$, and $\tilde{\rho}_{\Lambda}$, $\tilde{p}_{\Lambda}$ are given by Eqs. (\ref{eq7}, \ref{eq9}) respectively.\\

The evolution of the deceleration parameter is shown in Fig.1 for the parameter values: $\Omega_{m0}=0.27$, $\Omega_{r0}=0$, 
$\Omega_{\Lambda 0}=0.73$, $\omega_0=-1$ (which are consistent with current observations) and some values of $\beta$. Note that for $\beta=0.5,0.7$, the values of the transition redshift $z_T$ are consistent with the current observational data \cite{yin}, \cite{ruth}.

\noindent
\begin{center}
\includegraphics [scale=0.7]{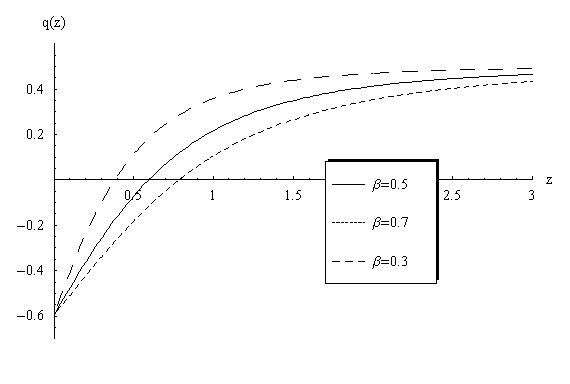}
\end{center}
\begin{center}
\it{Figure 1: Deceleration parameter versus redshift, considering $\omega_0=-1$, $\Omega_{m0}=0.27$, 
$\Omega_{r0}=0$, $\Omega_{\Lambda 0}=0.73$ and $\beta=0.3, 0.5, 0.7$.}
\end{center}
The evolution of the equation of state parameter $\omega=p_{\Lambda}/\rho_{\Lambda}$ is shown in fig.2 for $\beta=0.5$. It runs from nearly $0$ at high redshifts to $-1$ at $z->0$, behaving like some scalar-field models of dark energy \cite{copeland}.
\begin{center}
\includegraphics [scale=0.7]{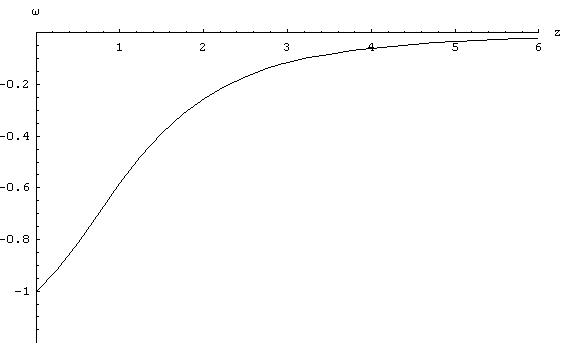}
\end{center}
\begin{center}
\it{Figure 2: Equation of state parameter versus redshift, considering $\omega_0=-1$, $\Omega_{m0}=0.27$, 
$\Omega_{r0}=0$, $\Omega_{\Lambda 0}=0.73$ and $\beta=0.5$.}
\end{center}

\section{Discussion}

We propose a new infrared cut-off for the holographic dark energy model, which includes a term proportional to $\dot{H}$. Contrary to the holographic dark energy based on the event horizon, this model depends on local quantities, avoiding in this way the causality problem. Calculating the contribution at radiation epoch to radiation
by dark energy from Eq. (\ref{eq6}), it follows that in order to be consistent with the Big-Bang nucleosintesis theory constraints, the constant $\beta$ should be very close to 0.5. At small redshift ($z<1$), note that in the expression of density for dark energy Eq.(\ref{eq7}), there are two terms which track dark matter and radiation, respectively. So this model avoids the coincidence problem.
The only parameter in this model which needs to be fitted by observational data is the new parameter $\beta$. Once $\beta$ is fixed by the appropriate value of the transition redshift $z_T$ (see Fig.1), the parameter $\alpha$ becomes fixed by Eq. (\ref{eq11}) (if we take $\beta\approx 0.5$, then $\alpha\approx 0.93$ and $z_T\approx 0.67$). From Fig.1 we see that the change from deceleration to acceleration takes place for data-consistent values of the model parameters, showing that this model is viable phenomenologically, although still to be studied the field model that justify the presence of $\dot{H}$ term in this kind of dark energy density.
 
\section*{Acknowledgments}
This work was supported by the Universidad del Valle, under project CI-7713.

\end{document}